\documentclass[english]{article}
\title{Computational Tutorial on Gr$\ddot{o}$bner bases embedding Sage in \LaTeX{} with \textsf{Sage\TeX}}
\author{Edinah K. Gnang}
\usepackage{hyperref}
\usepackage{sagetex}
\usepackage[T1]{fontenc}
\usepackage[latin9]{inputenc}
\usepackage{amsmath}
\usepackage{esint}
\usepackage{babel}
\usepackage{amssymb}

\begin{document}
\maketitle

\section{Introduction}
A comprehensive treatment of the theory of Gr$\ddot{o}$bner bases is far beyond the scope of this tutorial. The reader is reffered to 
\cite{Cox-Little-O'shea}, \cite{Buch}, \cite{Gathen-Gerhard} for a more detailed discussion. Gr$\ddot{o}$bner bases offers 
a powerful algorithmic criterion for the existence of solutions to a system of algebraic equations over algebraically closed 
fields. While the theory is rich and builds on the theory of Ideals and Varieties, the discussion in the persent tutorial will remain as elementary
as possible. The tutorial focuses on implementation aspects of Gr$\ddot{o}$bner bases computation via the Buchberger 
algorithm. The Sage\cite{sage} specific implementation discussed here is not meant to rival the efficient implementations 
available on Singular, Maxima and other Computer Algebra Systems(CAS). 
The tutorial merely attempts to provide an overview of the implementation details of 
the Buchberger algorithm. It must be pointed out that there are many other algorithms besides the Buchberger algorithm for computing 
Gr$\ddot{o}$bner bases, however for simplicity we restrict our attention to the Buchberger Algorithm.  
The reader can contact the author \footnote{http://www.mathematx.com}
, to obtain the \LaTeX{} source or to suggest possible corrections\\ 
The problem that the tutorial discusses is the determination of existence of solutions to a system of polynomial 
equations of the form.
\begin{equation}
f_1(x_1,\cdots,x_n) =\cdots= f_m(x_1,\cdots,x_n) = 0.
\end{equation}
where
\[
\forall \, 1 \le k \le m \quad f_k(x_1,\cdots,x_n)\in \mathbb{C}[x_1,\cdots,x_n].
\]
The determination of existence of solution can be established algorithmically via Gr$\ddot{o}$bner bases computation. 

\section{Initial Set up}
For simplicity we will be considering systems of polynomial equations with at most  
10 variables.\\

\begin{sageblock}
# Defining the variables
var('x0, x1, x2, x3, x4, x5, x6, x7, x8, x9')
# The prime assignment
P = Primes()
p0 = P.unrank(0); p1 = P.unrank(1); 
p2 = P.unrank(2); p3 = P.unrank(3); 
p4 = P.unrank(4); p5 = P.unrank(5); 
p6 = P.unrank(6); p7 = P.unrank(7); 
p8 = P.unrank(8); p9 = P.unrank(9);
\end{sageblock}

\section{Multivariate leading term}
The multivariate leading term functions determines the leading term of an 
input polynomial. 
The leading term for a given polynomial is identified using a monomial ordering.
We use in the tutorial a monomial ordering proposed by the author which follows 
from the fundamental theorem of arithmetics. The proposed ordering  has the 
advantage of considerably simplifying aspects of the implementation.

\begin{sageblock}
def multivariate_leading_term(f):
    # Expression used for specifying the 
    # type of the operation.
    add = x0+x1
    mul = x0*x1
    xpo = x0^x1
    cst = 2   
    if (f.operator() == add.operator()):
        # Collecting the terms
        L = f.operands()
        # Collecting the terms striped from their
        # coefficients
        L_strpd = list()
        for i in range(len(L)):
            if (L[i]).arguments() != ():
                if ((L[i]).operator() == mul.operator() or \
(L[i]).operator() == xpo.operator()):
                    cst_fctr = 1
                    lst_i = (L[i]).operands()
                    for j in range(len(lst_i)):
                        if (lst_i[j]).arguments() == ():
                            cst_fctr = lst_i[j]
                    L_strpd.append((expand(L[i]/cst_fctr),i))
                elif (L[i] == x0 or L[i] == x1 or L[i] == x2 or \
L[i] == x3 or L[i] == x4 or L[i] == x5 or L[i] == x6 or \
L[i] == x7 or L[i] == x8 or L[i] == x9 ):
                    L_strpd.append((L[i],i))
        # Storing the integer and the index associated with the term
        tmp_value = (L_strpd[0][0]).substitute(x0=p0,x1=p1,x2=p2,\
x3=p3,x4=p4,x5=p5,x6=p6,x7=p7,x8=p8,x9=p9)
        idx = L_strpd[0][1]
        for k in range(len(L_strpd)):
            if (L_strpd[k][0]).substitute(x0=p0,x1=p1,x2=p2,x3=p3,\
x4=p4,x5=p5,x6=p6,x7=p7,x8=p8,x9=p9) > tmp_value:
                tmp_value = (L_strpd[k][0]).substitute(x0=p0,x1=p1,\
x2=p2,x3=p3,x4=p4,x5=p5,x6=p6,x7=p7,x8=p8,x9=p9)
                idx = L_strpd[k][1]
        return L[idx]
    elif (f.operator() == mul.operator()):
        return f
    elif (f.operator() == xpo.operator()):
        return f 
    else :
        return f

\end{sageblock}

\section{The multivariate division function}
The multivariate division function is at the heart of the Buchberger algorithm for computing Gr$\ddot{o}$bner bases. 
This routine generalizes the familiar long division algorithm for single variable polynomials two non trivial ways.
First the algorithm extends the single variable long division algorithm to polynomials with multiple variables. Second the algorithm
allows for the division of single multivariate polynomial by a finite list of multivariate polynomials.
It goes without saying that the implementation relies quite heavily on the monomial ordering 
. We provide here both the pseudocode description of the algorithm as found in \cite{Cox-Little-O'shea} and our proposed Sage implementation.
\newpage
\begin{enumerate}
\item \textbf{Input:} $\left\{ f_{k}\right\} _{1\le k\le s},\, f$ 
\item \textbf{Output:} $\left\{ a_{k}\right\} _{1\le k\le s},\, r$
\item \textbf{Initialization}$\left\{ a_{k}\leftarrow0\right\} _{1\le k\le s}$;
$r\leftarrow0$ ; $p\leftarrow f$
\item \textbf{WHILE $p\ne0$ DO }

\begin{enumerate}
\item $i\leftarrow1$
\item divisionoccured $\leftarrow$ false
\item \textbf{WHILE $i\le s$ }divisionoccured $=$false \textbf{DO }

\begin{enumerate}
\item \textbf{IF $LT\left(f_{i}\right)$ }divides $LT\left(p\right)$ \textbf{THEN }

\begin{enumerate}
\item $a_{i}\leftarrow a_{i}+\frac{LT(p)}{LT(f_{i})}$
\item $p\leftarrow p-\left(\frac{LT(p)}{LT(f_{i})}\right)f_{i}$
\item divisionoccured $\leftarrow$ true
\end{enumerate}
\textbf{end}

\item \textbf{ELSE}

\begin{enumerate}
\item $i\leftarrow i+1$
\end{enumerate}
\textbf{end}

\end{enumerate}
\textbf{end}

\item \textbf{IF }divisionoccured $=$ \textbf{FALSE}

\begin{enumerate}
\item $r\leftarrow r+LT(p)$
\item $p\leftarrow p-LT(p)$
\end{enumerate}
\textbf{end}

\end{enumerate}
\textbf{end}

\item \textbf{End}
\end{enumerate}
\newpage
The implementation of the algorithm described above is also particularly insightfull.
\begin{sageblock}
def multivariate_division(f, List):
    # Initializing the output List
    L = list()
    for j in range(len(List)+1):
        L.append(0)
    # Initializing the Polynomial
    p = f
    while( p != 0 ):
        i = 0
        division_occured = 0
        while (i in range(len(List))) and (division_occured == 0):
            if List[i] == 0:
                i = i+1
            else:
                # Getting the Leading term of fi
                Lt_fi = multivariate_leading_term(List[i])
                #Getting the leading Monomial of fi
                Lm_fi = Lt_fi/Lt_fi.substitute(x0=1,x1=1,x2=1,x3=1,\
x4=1,x5=1,x6=1,x7=1,x8=1,x9=1)
                # Getting the Leading term of p
                Lt_p  = multivariate_leading_term(p)
                #Getting the leading Monomial of p
                Lm_p  = Lt_p/Lt_p.substitute(x0=1,x1=1,x2=1,x3=1,\
x4=1,x5=1,x6=1,x7=1,x8=1,x9=1)
                m_p  = Lm_p.substitute(x0=p0,x1=p1,x2=p2,x3=p3,\
x4=p4,x5=p5,x6=p6,x7=p7,x8=p8,x9=p9)
                m_fi = Lm_fi.substitute(x0=p0,x1=p1,x2=p2,x3=p3,\
x4=p4,x5=p5,x6=p6,x7=p7,x8=p8,x9=p9)
                if (gcd(m_p, m_fi) == m_fi or gcd(m_p, m_fi) == -m_fi):
                    L[i] = expand(L[i] + Lt_p/Lt_fi)
                    p = expand(p - List[i] * (Lt_p/Lt_fi))
                    division_occured = 1
                else :
                    i = i+1
        if (division_occured == 0):
            L[len(List)] = L[len(List)] + multivariate_leading_term(p)
            p = p - multivariate_leading_term(p)
    return L
\end{sageblock}

\newpage
\section{The multivariate least common multiple }
The multivariate least common multiple function determines the monomial least common multiple(LCM)
for an input pair of monomials. Our proposed monomial ordering reduces the monomial LCM computation to the familiar integer LCM computation.

\begin{sageblock}
def multivariate_monomial_lcm(t1, t2):
    # These 2 lines of code get rid of the coefficient of the leading terms
    m1 = t1/t1.substitute(x0=1,x1=1,x2=1,x3=1,x4=1,x5=1,x6=1,x7=1,x8=1,x9=1)
    m2 = t2/t2.substitute(x0=1,x1=1,x2=1,x3=1,x4=1,x5=1,x6=1,x7=1,x8=1,x9=1)
    
    # The following computes the lcm value associated with the monomial we seek
    monomial_lcm_value = lcm(m1.substitute(x0=p0,x1=p1,x2=p2,x3=p3,\
x4=p4,x5=p5,x6=p6,x7=p7,x8=p8,x9=p9), m2.substitute(x0=p0,x1=p1,x2=p2,\
x3=p3,x4=p4,x5=p5,x6=p6,x7=p7,x8=p8,x9=p9))
    
    # The next section of line of codes recovers the 
    # monomial in question from the computed lcm integer.
    prime_factors = factor(monomial_lcm_value)
    factor_list = list(prime_factors)
    m = 1
    for i in range(len(factor_list)):
        tmp_list = list(factor_list[i])
        if (tmp_list[0]==p0):
            m = m*x0^Integer(tmp_list[1])
        elif (tmp_list[0]==p1):
            m = m*x1^Integer(tmp_list[1])
        elif (tmp_list[0]==p2):
            m = m*x2^Integer(tmp_list[1])
        elif (tmp_list[0]==p3):
            m = m*x3^Integer(tmp_list[1])
        elif (tmp_list[0]==p4):
            m = m*x4^Integer(tmp_list[1])
        elif (tmp_list[0]==p5):
            m = m*x5^Integer(tmp_list[1])
        elif (tmp_list[0]==p6):
            m = m*x6^Integer(tmp_list[1])
        elif (tmp_list[0]==p7):
            m = m*x7^Integer(tmp_list[1])
        elif (tmp_list[0]==p8):
            m = m*x8^Integer(tmp_list[1])
        elif (tmp_list[0]==p9):
            m = m*x9^Integer(tmp_list[1])
    return m
\end{sageblock}

\section{The multivariate substraction polynomial}
The multivariate subtraction polynomial function is an important subroutine for the Buchberger Algorithm.
Its inner working are quite reminiscent of the elimination procedure in Gaussian 
elimination and it's implementation is just as straight forward.

\begin{sageblock}
def multivariate_S_polynomials(List):
    L = list()
    for i in range(len(List)-1):
        for j in range(i+1,len(List)):
            Lt_fi = multivariate_leading_term(List[i])
            Lt_fj = multivariate_leading_term(List[j])
            monomial_lcm = multivariate_monomial_lcm(Lt_fi, Lt_fj)
            Sij = expand(List[i]*(monomial_lcm/Lt_fi) - \
List[j]*(monomial_lcm/Lt_fj))
            L.append(Sij)
    return L
\end{sageblock}

\section{multivariate reduction}
The multivariate reduction function controls the bounds on the degree of the polynomials
to be adjoined to the Gr$\ddot{o}$bner bases. The reduction is achieved by dividing the 
polynomial by the greatest common divisor(GCD) of all the terms making up the polynomial. 
It also follows from our proposed ordering that the monomial GCD also reduces to the 
familiar integer GCD computation.
\newpage
\begin{sageblock}
def multivariate_reduce_polynomial(f):
    # Checks to see if there is a constant term 
    # in which case no reduction is needed
    if f.substitute(x0=0,x1=0,x2=0,x3=0,\
x4=0,x5=0,x6=0,x7=0,x8=0,x9=0) != 0 :
        return f
    elif f == 0:
        return f
    else:
        L = list(f.iterator())

        # The next piece of code determines i
        # if the expression is a monomial
        prd = 1
        for j in range(len(L)):
            prd = prd*L[j]
        if (prd == f):
            return 1
        elif ((len(L) == 2) and (L[0]^L[1] == f)):
            return f

        for i in range(len(L)):
            # The next line of code gets rid of 
            # the coefficients in the list
            L[i] = L[i]/(L[i].substitute(x0=1,x1=1,x2=1,x3=1,\
x4=1,x5=1,x6=1,x7=1,x8=1,x9=1))
            L[i] = Integer(L[i].substitute(x0=p0,x1=p1,x2=p2,\
x3=p3,x4=p4,x5=p5,x6=p6,x7=p7,x8=p8,x9=p9))
            
        # Computing the greatest common divisior
        cmn_fctr = gcd(L)

        if cmn_fctr == 1 :
            return f












        else :
            # The next section of line of codes recover the monomial
            # in question from the computed gcd integer.
            prime_factors = factor(cmn_fctr)
            factor_list = list(prime_factors)
            m = 1
            for i in range(len(factor_list)):
                tmp_list = list(factor_list[i])
                if (tmp_list[0]==p0):
                    m = m*x0^Integer(tmp_list[1])
                elif (tmp_list[0]==p1):
                    m = m*x1^Integer(tmp_list[1])
                elif (tmp_list[0]==p2):
                    m = m*x2^Integer(tmp_list[1])
                elif (tmp_list[0]==p3):
                    m = m*x3^Integer(tmp_list[1])
                elif (tmp_list[0]==p4):
                    m = m*x4^Integer(tmp_list[1])
                elif (tmp_list[0]==p5):
                    m = m*x5^Integer(tmp_list[1])
                elif (tmp_list[0]==p6):
                    m = m*x6^Integer(tmp_list[1])
                elif (tmp_list[0]==p7):
                    m = m*x7^Integer(tmp_list[1])
                elif (tmp_list[0]==p8):
                    m = m*x8^Integer(tmp_list[1])
                elif (tmp_list[0]==p9):
                    m = m*x9^Integer(tmp_list[1])
            g = expand(f/m)
            return g

\end{sageblock}

\newpage
\section{Implementation of the Buchberger Algorithm}
In summary the key components required for the implementation of the Buchberger Algorithm are:\\
- \emph{The multivariate leading term}\\
- \emph{The multivariate division}\\
- \emph{The multivariate least common multiple}\\
- \emph{The multivariate Substraction polynomial}.\\
\\
Let us put the pieces together to implement the Buchberger algorithm for computing 
Gr$\ddot{o}$bner bases.\\
A Gr$\ddot{o}$bner bases $G$ of an ideal $\mathcal{I}$ generated by 
the set of polynomials 
$\{f_k\}_{1\le k\le m}\subset \mathbb{C}[x_1,\cdots,x_{10}]$ over the field $\mathbb{C}$ is
characterised by any one of the following properties, stated relative to some monomial order:\\
\\
- The ideal given by the leading terms of polynomials in $\mathcal{I}$ is itself generated by the leading terms of the basis $G$;\\
- The leading term of any polynomial in $\mathcal{I}$ is divisible by the leading term of 
some polynomial in the basis $G$;\\
- multivariate division of any polynomial in the polynomial 
ring $\mathbb{C}[x_1,\cdots,x_{10}]$ by $G$ gives a unique remainder;\\
- multivariate division of any polynomial in the ideal $\mathcal{I}$ by $G$ gives $0$.\\
\\
We provide here both the pseudocode description of the algorithm as found in \cite{Cox-Little-O'shea} and our proposed Sage implementation.

\subsubsection*{The Main Theorem and the Buchberger Algorithm }

Let $\left\{ f_{k}\right\} _{1\le k\le t}\ne0$ be a polynomial ideal.
Then a Groebner basis for $\mathcal{I}$ can be constructed in finitely
steps.

Here is the \emph{Buchberger algorithm}: 
\begin{enumerate}
\item \textbf{Input:} $F=f_{1},\cdots,f_{s}$ 
\item \textbf{Output:} a Groebner basis $G=\left(g_{1}\cdots,g_{t}\right)$
for $\mathcal{I}$, with $F\subset G$ 
\item $G\leftarrow F$ 
\item \textbf{repeat}

\begin{enumerate}
\item $G^{\prime}\leftarrow G$ 
\item \textbf{for} each pair \{$p$,$q$\},$p\ne q\in G$ \textbf{do}

\begin{enumerate}
\item $S$$\leftarrow$$\overline{S\left(p,q\right)}^{G^{\prime}}$ 
\item \textbf{if $S\ne0$ then $G\leftarrow G\cup\left\{ S\right\} $} 
\end{enumerate}
\textbf{end}

\end{enumerate}
\textbf{until $G=G^{\prime}$}

\end{enumerate}
Where $\overline{S\left(p,q\right)}^{G^{\prime}}$ denotes the remainder
on division of $S\left(p,q\right)$ by the ordered set of elements
of $G^{\prime}$.

\subsubsection{Reduced Gr$\ddot{o}$bner bases}

A \emph{reduced Gr$\ddot{o}$bner bases} for a polynomial Ideal $\mathcal{I}$ is
a Gr$\ddot{o}$bner bases for $\mathcal{I}$ such that : 
\begin{enumerate}
\item $a_{deg(f)}=1$ for all $p\in G$ 
\item For all $p\in G$ no monomials of $p$ lies in $\left\{ \left\langle \boldsymbol{LT}\left(G-p\right)\right\rangle \right\} $ 
\end{enumerate}
where $\boldsymbol{LT}$ denote the leading term.

\subsubsection{Sage implementation of the Algorithm}
The initialization part which specifies the following system of algebraic equations
\[
f_1:=2x_0x_2+4x_1x_2-6  \, = \, f_2:=x_2^2-x_2 \,=\, f_3:=x_1^2-x_1 \,=\, f_4:=x_0^2-x_0 \, = \, 0
\]
\begin{sageblock}
# The Ideal of interest is generated 
# by the following polynomials
f1 = expand((x0+2*x1)*(2*x2)) - 6
f2 = x2^2-x2
f3 = x1^2-x1
f4 = x0^2-x0

# Generators of the Ideal
# for solving f1=f2=f3=f4=0
I = list()
I.append(f1)
I.append(f2)
I.append(f3)
I.append(f4)

# Initialization step 
I_curr = list()
for i in range(len(I)):
    I_curr.append(I[i])

l_old = 0
l_new = len(I_curr)
\end{sageblock}

\newpage

The main loop for the Algorithm corresponds to the following 

\begin{sageblock}
# Boolean variable tracking contradictions
finished = 0

while l_old != l_new and finished == 0:
    # Computes the single pass of the substraction polynomials
    S  = multivariate_S_polynomials(I_curr)
    print '\n\n The subtraction polynomials yield'
    for i in range(len(S)):
        S[i] = multivariate_reduce_polynomial(S[i])
        print 'S[',i,']= ',S[i]

    # The instruction bellow is the lazy way of getting rid of the duplicates.
    St = Set(S)
    S = list(St)

    #Recording the size of the Ideal generator set before the division
    l_old = len(I_curr)
    for i in range(len(S)):
        tmp_list = multivariate_division(S[i], I_curr)
        if tmp_list[len(tmp_list)-1]!=0:
            I_curr.append(tmp_list[len(tmp_list)-1])

    # Printing the result of the first pass of the Buchberger algorithm.
    print '\n\n The Current generator for the Ideal is given by'
    for i in range(len(I_curr)):
        print I_curr[i]
        if I_curr[i] == I_curr[i].substitute(x0=0,x1=0,x2=0,x3=0,\
x4=0,x5=0,x6=0,x7=0,x8=0,x9=0) and I_curr[i].substitute(x0=0,x1=0,\
x2=0,x3=0,x4=0,x5=0,x6=0,x7=0,x8=0,x9=0) != 0 :
            finished = 1

    #recording the size of the generator set after the division
    l_new = len(I_curr)

I = I_curr
\end{sageblock}

\newpage
\section{Running the Script}
For illustration we compute Gr$\ddot{o}$bner bases for the Ideal induced by the system of 
equations 
\[
f_1:=\sage{f1} \, = \, f_2:=\sage{f2} \,=\, f_3:=\sage{f3} \,=\, f_4:=\sage{f4} \, = \, 0
\]
The output of the computation is the list of size $\sage{len(I_curr)}$ which corresponds to
the sought after Gr$\ddot{o}$bner bases $G$.
\[G_0 = \sage{I[0]}\] 
\[G_1 = \sage{I[1]}\] 
\[G_2 = \sage{I[2]}\] 
\[G_3 = \sage{I[3]}\] 
\[G_4 = \sage{I[4]}\] 
\[G_5 = \sage{I[5]}\] 
\[G_6 = \sage{I[6]}\]
The criteria for existence of solution to a system of polynomial equations is therefore 
summarized as follows:\\
if the inputed polynomial admitted no common root then the Gr$\ddot{o}$bner bases would 
include the constant polynomial which constitutes a certificate for non existence of 
solution to the given system of algebraic equations.
\[
f_1(x_1,\cdots,x_n) =\cdots= f_m(x_1,\cdots,x_n) = 0.
\] 

\section*{Acknowledgments}

The author is grateful to Professor Doron Zeilberger, Professor
Henry Cohn, Professor Mario Szegedy, Professor Vladimir Retakh and
Professor Ahmed Elgammal for insightful discussions and suggestions.
The author was partially supported by the National Science Foundation 
grant NSF-DGE-0549115 and Microsoft Research New England .

\end{document}